\documentclass[10pt,journal,letterpaper,compsoc]{IEEEtran}

\ifCLASSOPTIONcompsoc
  \usepackage[nocompress]{cite}
\else
  \usepackage{cite}
\fi
\ifCLASSINFOpdf
  \usepackage[pdftex]{graphicx}
  \usepackage{epstopdf}
  \DeclareGraphicsExtensions{.pdf,.jpeg,.png,.eps}
\else
  \usepackage[dvips]{graphicx}
  \DeclareGraphicsExtensions{.eps}
\fi
\usepackage{array}

\usepackage{mdwtab}

\begin{document}
%
\title{A Survey of XML Tree Patterns}
%
%
%
%

\author{Marouane Hachicha and J\'{e}r\^{o}me Darmont, \IEEEmembership{Member,~IEEE~Computer~Society}
\IEEEcompsocitemizethanks{\IEEEcompsocthanksitem Marouane Hachicha and J\'{e}r\^{o}me Darmont are from the Universit\'{e} de Lyon (ERIC Lyon 2), 5 avenue Pierre Mend\`{e}s-France, 69676 Bron Cedex, France.\protect\\
E-mails: marouane.hachicha@univ-lyon2.fr, jerome.darmont@univ-lyon2.fr
}
\thanks{}}

%
%

\markboth{TKDE,~Vol.~X, No.~X, September~2011}%
{Hachicha \& Darmont: A Survey of XML Tree Patterns}
%


\IEEEcompsoctitleabstractindextext{%
\begin{abstract}
With XML becoming an ubiquitous language for data interoperability purposes in various domains, efficiently querying XML data is a critical issue. This has lead to the design of algebraic frameworks based on tree-shaped patterns akin to the tree-structured data model of XML. Tree patterns are graphic representations of queries over data trees. They are actually matched against an input data tree to answer a query. Since the turn of the twenty-first century, an astounding research effort has been focusing on tree pattern models and matching optimization (a primordial issue). This paper is a comprehensive survey of these topics, in which we outline and compare the various features of tree patterns. We also review and discuss the two main families of approaches for optimizing tree pattern matching, namely pattern tree minimization and holistic matching. We finally present actual tree pattern-based developments, to provide a global overview of this significant research topic.
\end{abstract}

}

\maketitle

\IEEEdisplaynotcompsoctitleabstractindextext

%
\IEEEpeerreviewmaketitle

%
\ifCLASSOPTIONcompsoc
  \noindent\raisebox{2\baselineskip}[0pt][0pt]%
  {\parbox{\columnwidth}{\section{Introduction}\label{sec:introduction}%
  \global\everypar=\everypar}}%
  \vspace{-1\baselineskip}\vspace{-\parskip}\par
\else
  \section{Introduction}\label{sec:introduction}\par
\fi
%

%
%
%
%


\IEEEPARstart{S}{ince} its inception in 1998, the eXtended Markup Language (XML) \cite{Quin06} has emerged as a standard for data representation and exchange over the Internet, as many (mostly scientific, but not only) communities adopted it for various purposes, e.g., mathematics with MathML \cite{mathml}, chemistry with CML \cite{cml}, geography with GML \cite{gml} and e-learning with SCORM~\cite{scorm}, just to name a few. As XML became ubiquitous, efficiently querying XML documents quickly appeared primordial and standard XML query languages were developed, namely XPath \cite{ClarkD99} and XQuery \cite{BoagCFFRS07}. Research initiatives also complemented these standards, to help fulfill user needs for XML interrogation, e.g., XML algebras such as TAX \cite{JagadishLST01} and XML information retrieval~\cite{TrotmanPL06}. 

Efficiently evaluating path expressions in a tree-structured data model such as XML's is crucial for the overall performance of any query engine~\cite{MichielsMS07}. Initial efforts that mapped XML documents into relational databases queried with SQL \cite{DeutschFS99,ShanmugasundaramGTZDN99} induced costly table joins. Thus, algebraic approaches based on tree-shaped patterns became popular for evaluating XML processing \emph{natively} instead \cite{PaparizosJ05,PaparizosJ06}. Tree algebras indeed provide a formal framework for query expression and optimization, in a way similar to relational algebra with respect to the SQL language \cite{ChamberlinB74}. 

\begin{figure}[hbt]
\centering
\includegraphics[width=8.45cm]{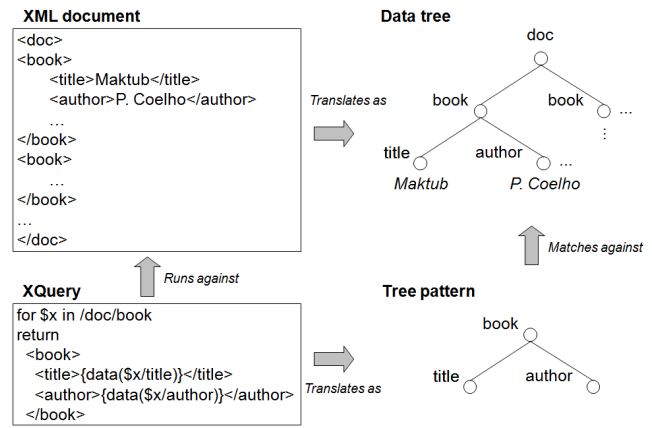}
\caption{Tree representation of XML documents and queries}
\label{fig-intro}
\end{figure}

In this context, a tree pattern (TP), also called pattern tree or tree pattern query (TPQ) in the literature, models a user query over a data tree. Simply put, a tree pattern is a graphic representation that provides an easy and intuitive way of specifying the interesting parts from an input data tree that must appear in query output. More formally, a TP is \emph{matched} against a tree-structured database to answer a query~\cite{AmerYahiaCLS01}. Figure~\ref{fig-intro} exemplifies this process. The upper left-hand side part of the figure features a simple XML document (a book catalog), and the lower left-hand side a sample XQuery that may be run against this document (and that returns the title and author of each book). The tree representations of the XML document and the associated query are featured on the upper and lower right-hand sides of Figure~\ref{fig-intro}, respectively. At the tree level, answering the query translates in matching the TP against the data tree. This process can be optimized and outputs a data tree that is eventually translated back as an XML document.

Since the early 2000s, a tremendous amount of research has been based on, focusing on, or exploiting TPs for various purposes. However, few related reviews exist. Gou and Chirkova extensively survey querying techniques over persistently stored XML data \cite{GouC07}. Although the intersection between their paper and ours is not empty, both papers are complementary. We do not address approaches related to the relational storage of XML data. By focusing on native XML query processing, we complement Gou and Chirkova's work with specificities such as TP structure, minimization approaches and sample applications. Moreover, we cover the many matching optimization techniques that have appeared since 2007. Other recent surveys are much shorter and focus on a particular issue, i.e., twig queries \cite{Lakshmanan09} and holistic matching \cite{GrimsmoB10}. 
 
The aim of this paper is thus to provide a global and synthetic overview of more than ten years of research about TPs and closely related issues. For this sake, we first formally define TPs and related concepts in Section~\ref{background}. Then, we present and discuss various alternative TP structures in Section~\ref{sec:PatternTreeStructures}. Since the efficiency of TP matching against tree-structured data is central in TP usage, we review the two main families of TP matching optimization methods (namely, TP minimization and holistic matching approaches), as well as tangential but nonetheless interesting methods, in Section~\ref{sec:PatternTreeMatchingOptimization}. Finally, we briefly illustrate the use of TPs through actual TP-based developments in Section~\ref{sec:PatternTreeUsages}. We conclude this paper and provide insight about future TP-related research in Section~\ref{conclusion}.

\section{Background}
\label{background}

\IEEEPARstart{I}{n} this section, we first formally define all the concepts used in this paper (Section~\ref{sec:Definitions}). We also introduce a running example that illustrates throughout the paper how the TPs and related algorithms we survey operate (Section~\ref{sec:RunningExample}).

\subsection{Definitions}
\label{sec:Definitions}

\subsubsection{XML document}

XML is known to be a simple and very flexible text format. It is essentially employed to store and transfer text-type data. The content of an XML document is encapsulated within elements that are defined by tags \cite{Quin06}. These elements can be seen as a hierarchy organized in a tree-like structure.

\subsubsection{XML fragment}

An XML document may be considered as an ordered set of elements. Any element may contain subelements that may in turn contain subelements, etc. One particular element, which contains all the others, is the document's root. Any element (and its contents) different from the root is termed an XML fragment. An XML fragment may be modeled as a finite rooted, labeled and ordered tree.

\subsubsection{Data tree}

An XML document (or fragment) may be modeled as a data tree $t = (r, N, E)$, where $N$ is a set of nodes, $r \in N$ is the root of $t$, and $E$ is a set of edges stitching together couples of nodes $(n_i, n_j) \in N \times N$.

\subsubsection{Data tree collection}

An XML document considered as a set of fragments may be modeled as a data tree collection (also named forest in TAX~\cite{JagadishLST01}), which is itself a data tree.

\subsubsection{Data subtree}

Given a data tree $t = (r, N, E)$, $t^{'} = (r^{'}, N^{'}, E^{'})$ is a subtree of $t$ if and only if:

\begin{enumerate}
	\item $N^{'} \subseteq N$;
	\item let $e^{'} \in E^{'}$ be an edge connecting two nodes $(n_i^{'}, n_j^{'})$ of $t^{'}$; there exists an edge $e \in E$ connecting two nodes $(n_i, n_j)$ of $t$ such that $n_i = n_i^{'}$ and $n_j = n_j^{'}$.
\end{enumerate}

\subsubsection{Tree pattern} 

A tree pattern $p$ is a pair $p = (t, F)$ where:

\begin{enumerate}
	\item $t = (r, N, E)$ is a data tree;
	\item $F$ is a formula that specifies constraints on $t$'s nodes. 
\end{enumerate}

Basically, a TP captures a useful fragment of XPath~\cite{FlescaFM03}. Thus, it may be seen as the translation of a user query~\cite{ChenJLP03}. Translating an XML query plan into a TP is not a simple operation. Some XQueries are written with complex combinations of XPath and FLWOR expressions, and imply more than one TP. Thus, such queries must be broken up into several TPs. Only a single XPath expression can be translated into a single TP. The more complex a query is, the more its translation into TP(s) is difficult~\cite{MichielsMS07}. Starting from TPs to express user queries in a first stage, and optimize them in a second stage is a very effective solution used in XML query optimization~\cite{ChenJLP03}.

\subsubsection{Tree pattern matching}
\label{matching}

Matching a TP $p$ against a data tree $t$ is a function $f: p \rightarrow t$ that maps nodes of $p$ to nodes of $t$ such that:

\begin{itemize}
	\item structural relationships are preserved, i.e., if nodes $(x, y)$ are related in $p$ through a parent-child relationship, denoted PC for short or by a simple edge / in XPath (respectively, an ancestor-descendant relationship, denoted AD for short or by a double edge // in XPath), their counterparts $(f(x), f(y))$ in $t$ must be related through a PC (respectively, an AD) relationship too;
	\item formula $F$ of $p$ is satisfied. 
\end{itemize}

The output of matching a TP against a data tree is termed a witness trees in TAX~\cite{JagadishLST01}.

\subsubsection{Tree pattern embedding}

Embedding a TP $p$ into a data tree $t$ is a function $g: p \rightarrow t$ that maps each node of $p$ to nodes of $t$ such that structural relationships (PC and AD) are preserved. 

The difference between embedding and matching is that embedding maps a TP against a data tree's \emph{structure} only, whereas matching maps a TP against a data tree's structure \emph{and contents}~\cite{LakshmananRWZ04}. In the remainder of this paper, we use the more general term matching when referring to mapping TPs against data trees.

\begin{figure*}[hbt]
\centering
\includegraphics[width=16cm]{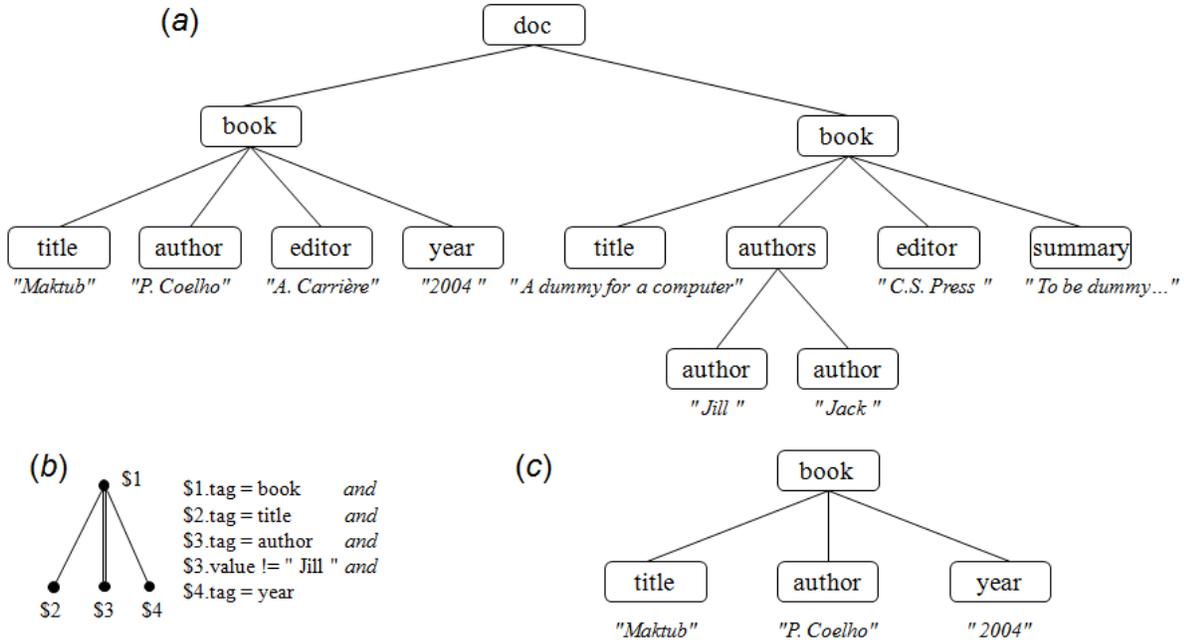} 
\caption{Sample data tree (a), tree pattern (b) and witness tree (c)}
\label{fig-runex}
\end{figure*}

\subsubsection{Boolean tree pattern}

A Boolean TP $b$ yields a ``true'' (respectively ``false'') result when successfully (respectively unsuccessfully) matched against a data tree $t$. In other words, $b$ has no output node(s)~\cite{WangYL09}: it yields a ``true'' result when $t$ \emph{structurally} matches (embeds) $p$.

\subsection{Running example}
\label{sec:RunningExample}

Let us consider the data tree from Figure~\ref{fig-runex}(a), which represents a collection of books. The root \textit{doc} gathers \textit{books}, each described by their \textit{title}, \textit{author(s)}, \textit{editor}, \textit{year} of publication and \textit{summary}. Data tree nodes are connected by simple edges (/), i.e., PC relationships. Books do not necessarily bear the same structure. For instance, a summary is not present in all books, and some books are written by more than one author.

The TP from Figure~\ref{fig-runex}(b) selects book titles, authors, and years. Moreover, formula $F$ indicates that author must be different from Jill. Generally, a formula is a boolean combination of predicates on node types (e.g., $\$1.tag =$ book) and/or values (e.g., $\$3.value~!=$~``Jill''). Matching this TP against the data tree from Figure~\ref{fig-runex}(a) outputs the data tree (or witness tree) from Figure~\ref{fig-runex}(c). Only one book is selected from Figure~\ref{fig-runex}(a), since the other one ($title = $~``A dummy for a computer''):
\begin{enumerate}
	\item does not contain a year element;
	\item is written by author Jill, which contradicts formula $F$.
\end{enumerate}

Finally, the AD relationship $\$1//\$3$ in Figure~\ref{fig-runex}(b)'s TP is correctly taken into account. 
It is not the structure of the book element with $title =$ ``A dummy for a computer'' that disqualifies it, but the fact that one of its authors is Jill. If this author was Gill, the book would be output.

\section{Tree pattern structures}
\label{sec:PatternTreeStructures}

\IEEEPARstart{W}{e} review in this section the various TP structures found in the literature. Most have been proposed to support XML algebras (Section~\ref{Pattern-xml-algebras}), but some have also been introduced for specific optimization purposes (Section~\ref{Pattern-trees-used-in-optimisation}). We conclude this section by discussing their features in Section~\ref{Discussion}.

\subsection{Tree patterns in algebraic frameworks}
\label{Pattern-xml-algebras}

The first XML algebras have appeared in 1999~\cite{beech99,BeeriT99} in conjunction with efforts aiming to define a powerful XML query language~\cite{ChamberlinRF-quilt00,DeutschFFLS98,IshikawaKKN99,MeccaMA98,ShinagawaKI00}. Note that they have appeared before the first specification of XQuery, the now standard XML query language, which was issued in 2001~\cite{BoagCFFRS07}. The aim of an XML tree algebra is to feature a set of operators to manipulate and query data trees. Query results are also data trees.

\subsubsection{TAX tree pattern}
\label{Pattern-tree-of-TAX-XML-tree-algebra}

The Tree Algebra for XML (TAX) \cite{JagadishLST01} is one of the most popular XML algebras. TAX's TP preserves PC and AD relationships from an input ordered data tree in output, and satisfies a formula that is a boolean combination of predicates applicable to nodes. The example from Figure~\ref{fig-runex}(b) corresponds to a TAX TP, save that node relationships are simple edges labeled AD or PC in TAX instead of being expressed as single or double edges. The TAX TP is the most basic TP used in algebraic contexts. It has thus been greatly extended and enhanced.

\subsubsection{Generalized tree pattern}
\label{GTPs:-Generalized-Tree-Patterns}

The idea behind generalized tree patterns (GTPs) is to associate more options with TP edges in order to enrich matching. In TAX, one absent pattern node in the matched subtree prevents it to appear in output. A GTP extends the classical TAX TP by creating groups of nodes to facilitate their manipulation, and by enriching edges to be extracted by the \textit{mandatory/optional} matching option~\cite{ChenJLP03}. 

\begin{figure}[hbt]
\centering
\includegraphics[width=5cm]{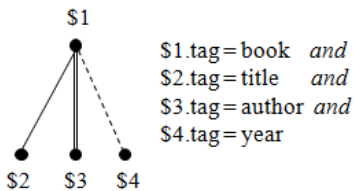}
\caption{Sample generalized tree pattern}
\label{fig-gtp}
\end{figure}

Figure~\ref{fig-gtp} shows an example of GTP, where the edge connecting the year element to its parent book node is dotted (i.e., optional), and title and author nodes are connected to the same parent book node with solid (i.e., mandatory) edges. This GTP permits to match all books described by their title and author(s), mandatorily, and that may be described by their year of publication. Matching it against the data tree from Figure~\ref{fig-runex}(a) outputs both books (``Maktub'' and ``A dummy for a computer''), even though the second one does not contain a \textit{year} element, while ensuring that the \emph{title} and \emph{author} elements exist in the matched book subtrees. 

Finally, note that the GTP from Figure~\ref{fig-gtp}, unlike the TP from Figure~\ref{fig-runex}(b), does not include an $author$ != ''Jill'' clause in its formula. Retaining this predicate would not allow the second book ($title$ = ``A dummy for a computer'') to be matched by this GTP, since the boolean combination of formula elements (related with \textit{and}) cannot be verified.

\subsubsection{Annotated tree pattern}
\label{APTs:-Annotated-Pattern-Trees}

A feature, more than a limitation, of the TAX TP is that a set of subelements from the input data tree may all appear in the output data tree. For example, a TP with a single \emph{author} node can match against a \emph{book} subtree containing several \emph{author} subelements. Annotated pattern trees (APTs) from the Tree Logical Class (TLC) algebra~\cite{PaparizosWLJ04} solve this problem by associating matching specifications to tree edges. Matching options are: 

\begin{itemize}
	\item    \texttt{+}: one to many matches;
	\item    \texttt{-}: one match only; 
	\item    \texttt{*}: zero to many matches;
	\item    \texttt{?}: zero or one match.
\end{itemize}       
                         
\begin{figure}[hbt]
\centering
\includegraphics[width=5.5cm]{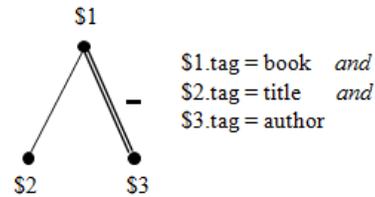}
\caption{Sample annotated pattern tree}
\label{fig-apt}
\end{figure}

Figure~\ref{fig-apt} shows an example of APT where the \texttt{-} option is employed to extract books written by one author only. This APT matches only the first book from Figure~\ref{fig-runex}(a) ($title =$ ``Maktub''), since the second has two authors. 

\subsubsection{Ordered annotated pattern tree}
\label{Extending-APTs:-Annotated-Pattern-Trees}

An APT, as a basic TAX TP, preserves the order output of nodes from the input data tree, whatever node order is specified in the TP. In other words, node order in witness trees is always the same as that of the input data tree. No reordering option is available, though it is essential when optimizing queries. To circumvent this problem, the APT used in TLC's \texttt{Select} and \texttt{Join} operators is supplemented with an order parameter (\textit{ord}) based on an order specification (\textit{O-Spec})~\cite{PaparizosJ05}. Four cases may occur:
\begin{enumerate}
	\item  \textit{empty:} output node order is unspecified; \textit{O-Spec} is empty;
	\item  \textit{maintain:} input node order is retained; \textit{O-Spec} duplicates input node order;
	\item  \textit{list-resort:} nodes are wholly sorted with respect to \textit{O-Spec}; input node order is forsaken; 
  	\item  \textit{list-add:} nodes are partially sorted with respect to \textit{O-Spec}.
\end{enumerate}

For example, associating the \textit{O-Spec} specification \emph{[author, title]} to the APT from Figure~\ref{fig-apt} permits to select books ordered by author first, and then title. Graphically, the \emph{author} node would simply appear on the left hand side of the witness tree and the \emph{title} node on the right hand side.

\subsection{Tree patterns used in optimization processes}
\label{Pattern-trees-used-in-optimisation}

Many TP matching optimization approaches extend the basic TP (Figure~\ref{fig-runex}(b)) to allow a broader range of queries. In this section, we survey the TPs that introduce new, interesting features with respect to those already presented in Section~\ref{Pattern-xml-algebras}.

\begin{figure*}[hbt]
\centering
\includegraphics[width=14.9cm]{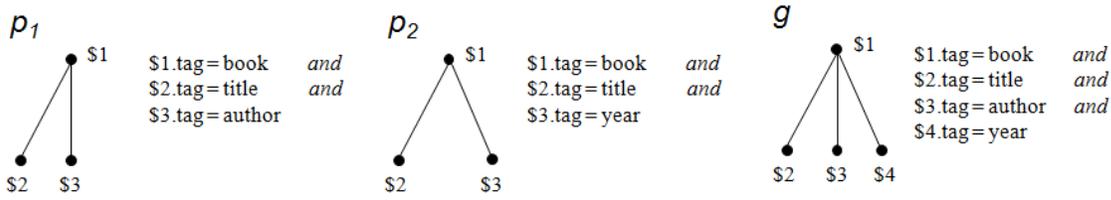}
\caption{Sample construction of global query pattern tree}
\label{fig-gqtp}
\end{figure*}

\subsubsection{Global query pattern tree}
\label{GQPT}

A global query pattern tree (G-QPT) is constructed from a set of possible ordered TPs proposed for the same query~\cite{Yichen04}. First, a root is created for the G-QPT. Then, each TP is merged with the G-QPT as follows: 
\begin{itemize}
	\item the TP root is merged with the G-QPT root;
	\item TP nodes are merged with G-QPT nodes with respect to node ordering and PC-AD relationships. 
\end{itemize}

For example, TPs $p_1$ and $p_2$ from Figure~\ref{fig-gqtp} together answer query ``select title, authors and year of publication of books'' and merge into G-QPT $g$.


\subsubsection{Twig pattern}
\label{Twig-Patterns}

\begin{figure}[hbt]
\centering
\includegraphics[width=5cm]{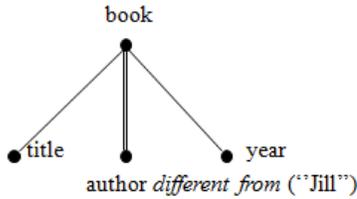}
\caption{Sample twig pattern}
\label{fig-twig}
\end{figure}

Twig patterns are designed for directed and rooted \emph{graphs}~\cite{KimelfeldS06}. Here, XML documents are considered having a graph structure, thanks to ID references connecting nodes. Twig patterns respect PC-AD node relationships and express node constraints in formula $F$ (called twig condition here) with operators such as \textit{equals to}, \textit{contains}, $\geq$, etc. In Figure~\ref{fig-twig}, we represent the twig pattern that is equivalent to the TP from Figure~\ref{fig-runex}(b). 

Moreover, distance-bounding twigs (DBTwigs) extend twig patterns to limit the large number of answers resulting from matching a classical twig pattern against a graph~\cite{KimelfeldS06}. This method is called Filtering + Ranking. Its goal is to filter data to be matched by indicating the length of paths corresponding to descendant edges. DBTwigs also permit to indicate the number (0, 1...) of nodes to be matched. All these parameters are indicated in the graphical representation of DBTwigs.

\subsubsection{Logical operator nodes}
\label{A-PT-IZADI}

\begin{figure}[hbt]
\centering
\includegraphics[width=4cm]{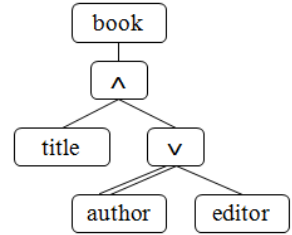}
\caption{Sample tree pattern with logical operator nodes}
\label{fig-logical}
\end{figure}

Izadi \textit{et al.} include in their formal definition of TPs a set $O$ of logical operator nodes~\cite{IzadiHH09}. $\wedge$, $\vee$ and $\oplus$ represent the binary AND, OR, and XOR logical operators, respectively. $\neg$ is the unary NOT operator. These operator nodes go further than GTPs' \emph{mandatory/optional} edge annotations by specifying logical relationships between their subnodes. For instance, Figure~\ref{fig-logical} features a TP that selects book titles \emph{and} either a set of authors \emph{or} an editor.

\subsubsection{Node degree and output node specification}
\label{Associating-options-to-PT-nodes}

In a TP, the degree of a tree node $x$, denoted $degree(x)$, represents its number of children~\cite{MiklauS02}. Miklau and Suciu define a maximal value for $degree(x)$ within a path expression optimization algorithm. This approach is mainly designed to check containment and equivalence of XPath fragments (Section~\ref{sec:EquivalenceAndContainmentTests}). Then, translating XPath expressions into TPs requires the identification of an output node marked with a wildcard ($\ast$) in the TP. For example, suppose that we are only interested in book titles from Figure~\ref{fig-runex}(a). Then, we would simply associate a $\ast$ symbol with the \emph{title} node in the TP from Figure~\ref{fig-runex}(b). Note that the other nodes must remain in the TP if they appear in formula $F$ and thus help specify the output node.

\subsubsection{Extended formula}
\label{sec:ExtendedFormula}

Lakshmanan~\textit{et al.} further specify how formula $F$ is expressed in a TP~\cite{LakshmananRWZ04}. They define $F$ as a combination of tag constraints (TCs), value-based constraints (VBCs) and node identity constraints (NICs). TCs specify constraints on tag names, e.g., $node.tag =$ ``book''. VBCs specify selection and join constraints on tag attributes and values using the operators $=$, $\neq$, $\leq$, $\geq$, $>$ and $<$. NICs finally determine whether two TP nodes are the same ($\approx$) or not. In addition to TCs, VBCs and NICs, wildcards ($\ast$) are associated with untagged nodes.

\begin{figure}[hbt]
\centering
\includegraphics[width=6cm]{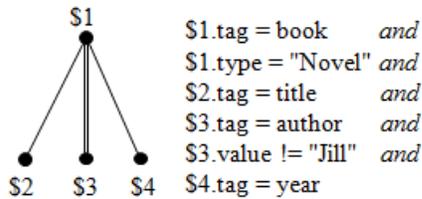}
\caption{Sample tree pattern with a value-based constraint}
\label{fig-constraints}
\end{figure}

For example, in Figure~\ref{fig-constraints}, we extend the TP from Figure~\ref{fig-runex}(b) with a VBC indicating that books to be selected must be of type ``Novel''. We suppose here that a type is associated with each \textit{book} node, that the first book ($title =$ ``Maktub'') is of type ``Novel'', and that the second book ($title =$ ``A dummy for a computer'') is of type ``Technical book''. Then, the output tree would of course only include the book entitled ``Maktub''.

\subsubsection{Extended tree pattern}
\label{extp}

\begin{figure}[hbt]
\centering
\includegraphics[width=4.5cm]{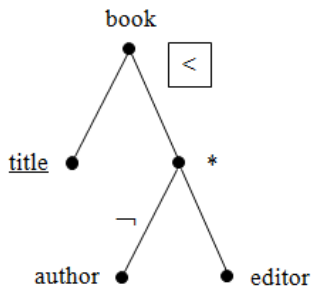}
\caption{Sample extended tree pattern}
\label{fig-extp}
\end{figure}

Extended TPs complement classical TPs with a negation function, wildcards and an order restriction \cite{LuLBCW-TKDE10}. For example, in Figure~\ref{fig-extp}, the negative function (denoted $\neg$) helps specify that we look for an edited book, i.e., with no author node. The wildcard node $\ast$ can match any single node in a data tree. Note that the wildcard has a different meaning here than in Section~\ref{Associating-options-to-PT-nodes}, where it denotes an output node, while output node names are underlined in extended TPs. Finally, the order restriction denoted by a $<$ in a box means that children of node book are ordered, i.e., title must come before the $\ast$ node.

\subsection{Discussion}
\label{Discussion}

In this section, we discuss and compare the TPs surveyed in Sections~\ref{Pattern-xml-algebras} and \ref{Pattern-trees-used-in-optimisation}. For this sake, we introduce four comparison criteria.

\begin{itemize}

	\item \emph{Matching power:} Matching encompasses two dimensions. Structural matching guarantees that only subtrees of the input data tree that map the TP are output. Matching by value is verifying formula $F$. We mean by matching power all the matching options (edge annotations, logical operator nodes, formula extensions...) beyond these basics. Improving matching power helps filter data more precisely.
	
	\item \emph{Node reordering capability:} Order is important in XML querying, thus modern TPs should be able to alter it~\cite{PaparizosJ05}. We mean by node reordering capability the ability of a TP to modify output node order when matching against any data tree. Note that node reordering could be classified as a matching capability, but the importance of ordering witness trees leads us to consider it separately. 
	
	\item \emph{Expressiveness:} Expressiveness states how a logical TP, i.e., a TP under its schematic form, translates into the corresponding physical implementation. The physical form of a TP may be either expressed with an XML query language (XQuery, generally), or an implementation within an XML database offering query possibilities. Using a TP for other purposes but querying, e.g., in optimization algorithms, is not accounted toward expressiveness. Note that a TP that cannot be expressed in physical form is usually considered useless.
	
	\item \emph{Supported optimizations:} TPs are an essential element of XML querying~\cite{AmerYahiaCLS01, AmerYahiaCS02}. Hence, many optimization approaches translate XML queries into TPs, optimize them, and then translate them back into optimized queries. Optimizing a TP increases its matching power. This criterion references the different kinds of optimizations supported by a given TP. 
	
\end{itemize}

TP comparison with respect to matching power, node reordering capability, expressiveness and supported optimizations follows (Sections~\ref{sec:MatchingPower}, \ref{sec:NodeReorderingCapability}, \ref{sec:Expressiveness} and \ref{sec:SupportedOptimizations}, respectively), and is synthesized in Section~\ref{sec:Synthesis}.

\subsubsection{Matching power}
\label{sec:MatchingPower}

The most basic TP is TAX's \cite{JagadishLST01}. Thus, matching a TAX TP against a data tree collection is very simple: it is only based on tree structure and node values specified in formula $F$. If $F$ is absent, a TAX TP matches all possible subtrees from the input data tree.

Associating a \textit{mandatory}/\textit{optional} status to GTP edges~\cite{ChenJLP03} increases the number of matched subtrees in output. Only one absent edge in a TAX TP with respect to a subtree containing all the other edges in this TP prevents matching, while the candidate subtree is ``almost perfect''. If this edge is labeled as \textit{optional} in a GTP, matching becomes successful. Even more flexibility is achieved by logical operator nodes~\cite{IzadiHH09}, which allow powerful matching options, especially when logical operators are combined.

With APTs, matching precision is further improved through edge annotations~\cite{PaparizosWLJ04}. Annotations in APTs force the matching process to extract data according to their nature. For example, a classical TAX TP such as the one from Figure~\ref{fig-runex}(b), extracting books by titles and authors, does not allow controlling the number of authors. APTs help select only books with, e.g., more than one author, as in the example from Figure~\ref{fig-apt}. Unfortunately, annotations only allow two maximum cardinalities: one or several. DBTwigs~\cite{KimelfeldS06} complement APTs in this respect, by allowing to choose the exact number of nodes to be matched. Node degree (number of children)~\cite{MiklauS02} or the negation function of extended TPs \cite{LuLBCW-TKDE10} can also be exploited for this sake.

Finally, up to now, we focused on TP structure because few differences exist in formulas. Graphically, twig patterns associate constraints with nodes directly in the schema \cite{KimelfeldS06} while in the TAX TP and its derivatives, they appear in formula $F$, but this is purely cosmetic. Only TCs, VBCs and NICs help further structure formula $F$ \cite{LakshmananRWZ04}.

\subsubsection{Node reordering capability}
\label{sec:NodeReorderingCapability}

Despite many studies model XML documents as unordered data trees, order \emph{is} essential to XML querying~\cite{PaparizosJ05}. In XQuery, users can (re)order output nodes simply through the \texttt{Order by} clause inherited from SQL \cite{ChamberlinB74}. Hence, modern TPs should include ordering features~\cite{AmerYahiaCLS01, AmerYahiaCS02}.

However, among TAX TP derivatives, only ordered APTs and extended TPs feature reordering features. In G-QTPs, preorders associated with ordered TP nodes help determine output order~\cite{Yichen04}. Unfortunately, order is disregarded in all other TP proposals.

\subsubsection{Expressiveness}
\label{sec:Expressiveness}

TAX TPs and their derivatives (GTPs and APTs) do not translate into an XML query language, but they are implemented, through the TLC physical algebra \cite{PaparizosWLJ04}, in the TIMBER XML database management system \cite{PaparizosACJLNPSWWY03}. TIMBER permits to store XML in native format and offers a query interface supporting both classical XQuery fragments and TAX operators. Note that TAX operators include a \texttt{Group by} construct that has no equivalent in XQuery. 

Translating TAX TPs for XML querying follows nine steps:
\begin{enumerate}
	\item identify all TP elements in the \texttt{FOR} clause;
	\item push formula $F$'s predicates into the \texttt{WHERE} clause;
	\item eliminate duplicates with the help of the \texttt{DISTINCT} keyword;
	\item evaluate aggregate expressions in \texttt{LET} clauses;
	\item indicate tree variables to be joined (join conditions) via the \texttt{WHERE} clause; 
	\item enforce any remaining constraint in the \texttt{WHERE} clause; 
	\item evaluate \texttt{RETURN} aggregates; 
	\item order output nodes with the help of the \texttt{ORDER BY} clause; 
	\item project on the elements indicated in the \texttt{RETURN} clause. 
\end{enumerate}

Similarly, Lakshmanan~\textit{et al.} test the satisfiability of TPs translated from XPath expressions and XQueries, and then express them back in XQuery and evaluate them within the XQEngine XQuery engine \cite{XQEngine}. The other TPs we survey are used in various algorithms (containment and equivalence testing, TP rewriting, frequent TP mining...). Hence, their expressiveness is not assessed.

\subsubsection{Supported optimizations}
\label{sec:SupportedOptimizations}

Approaches purposely proposing TPs (i.e., algebraic approaches) are much fewer than approaches using TPs to optimize XML querying. Moreover, even algebraic approaches such as TAX do address optimization issues. Since XML queries are easy to model with TPs, researchers casually translate any XML query in TPs, which are then optimized and implemented in algorithms, XML queries or any other optimization framework. 

Hence, the biggest interest of the TP community lies in enriching and optimizing matching. Matching opportunities offered by TAX TPs, optional edges of GTPs, annotations, ordering specification and duplicate elimination of APTs, and extended TPs aim to achieve more results and/or better precision. GTP and APT matching characteristics prove their efficiency in TIMBER~\cite{PaparizosACJLNPSWWY03}.


Finally, satisfiability is an issue related to containment, a concept used in minimization approaches. Satisfiability means that there exists a database, consistent with the schema if one is available, on which the user query, represented by a TP, has a non-empty answer~\cite{Hidders03,LakshmananRWZ04,David08,BenediktFG08}.


\subsubsection{Synthesis}
\label{sec:Synthesis}

We recapitulate in Table~\ref{tab-synth} the characteristics of all TPs surveyed in this paper with respect to our comparison criteria.

\begin{table*}[hbt]
\renewcommand{\arraystretch}{1.3}
\caption{Synthesis of tree pattern characteristics}
\label{tab-synth}
\centering
\begin{tabular}{|l|c|c|c|c|}
\hline
\textbf{TP} & \textbf{Matching features} & \textbf{Reordering} & \textbf{Expressiveness} & \textbf{Supported optimizations} \\
\hline
\textbf{TAX TP} \cite{JagadishLST01} & Basic & No & TIMBER & Matching \\
\hline
\textbf{GTP} \cite{ChenJLP03} & Mandatory/optional edges & No & TIMBER & Matching \\
\hline
\textbf{APT} \cite{PaparizosWLJ04} & Edge cardinality & No & TIMBER & Matching \\
\hline
\textbf{Ordered APT} \cite{PaparizosJ05} & Order specification & Yes & TIMBER & Matching \\
\hline
\textbf{G-QPT} \cite{Yichen04} & Set of TPs & Yes & N/A & Unspecified \\
\hline
\textbf{Twig pattern} \cite{KimelfeldS06} & Graph structure & No & N/A & XML graph querying \\
\hline
\textbf{DBTwig} \cite{KimelfeldS06} & Filtering + Ranking & No & N/A & XML graph querying \\
& + Matching cardinality & & & \\
\hline
\textbf{Izadi} & Logical operator nodes & No & N/A & Matching \\
\textbf{\emph{et al.}'s TP} \cite{IzadiHH09}& + Potential target node & & & \\
\hline
\textbf{Miklau and} & Node degree & No & N/A & Containment and \\
\textbf{Suciu's TP} \cite{MiklauS02} & + Output node & & & equivalence testing \\
\hline
\textbf{Lakshmanan} & Tag constraints & & & \\
\textbf{\emph{et al.}'s TP} \cite{LakshmananRWZ04} & + Value-based constraints & No & XQuery & Satisfiability testing \\
& + Node identity constraints & & & \\
\hline
\textbf{Extended TP} & Negation function & Yes & N/A & Matching \\
\cite{LuLBCW-TKDE10} & + Wildcards & & & \\
\hline
\end{tabular}
\end{table*}

In summary, matching options in TPs are numerous and it would probably not be easy to set up a metric to rank them. The best choice, then, seems to select a TP variant that is adapted to the user's or designer's needs. However, designing a TP that pulls together most of the options we survey is certainly desirable to maximize matching power.

With respect to output node ordering, we consider the \textit{ord} order parameter introduced in APTs the simplest and most efficient approach. A list of ordered elements can indeed be associated with any TP, whatever the nature of the input data tree (ordered or unordered). However, it would be interesting to generalize the \textit{ord} specification to other, possibly more complex, operators beside \texttt{Select} and \texttt{Join}, the two only operators benefiting from ordering in APTs. 

Expressiveness is a complex issue. Translating XML queries into TPs is indeed easier than translating TPs back into an XML query plan. XQuery, although the standard XML query language, suffers from limitations such as the lack of a \texttt{Group by} construct. Thus, it is more efficient to implement TPs and exploit them to enrich XML querying in an ad-hoc environment such as TIMBER's. We think that the richer the pattern, with matching options, ordering specifications, possibility to associate with many operators (and other options if possible), the more efficient querying is, in terms of user need satisfaction. 

Finally, minimization, relaxation, containment, equivalence and satisfiability issues lead the TP community to optimize these tasks. However, most TPs used in this context are basic, unlike TPs targeted at matching optimization, which allows optimizing XML queries wherein are translated optimized TPs. In short, the best-optimized TP must be minimal, satisfiable, and offer as many matching options as possible.

\section{Tree pattern matching optimization}
\label{sec:PatternTreeMatchingOptimization}

\IEEEPARstart{T}{he} aim of TPs is not only to provide a graphical representation of queries over tree-structured data, but also and primarily, to allow matching queries against data trees. Hence, properly optimizing matching is primordial to achieve good query response time. 

In this section, we present the two main families of approaches for optimizing matching, namely minimization methods (Section~\ref{sec:PatternTreeMinimization}) and holistic matching approaches (Section~\ref{sec:holistic}). We also survey a couple of other specific approaches (Section~\ref{sec:SpecificApproaches}), before globally discussing and comparing all TP matching optimization methods (Section~\ref{sec:Discussion2}).

\subsection{Tree pattern minimization}
\label{sec:PatternTreeMinimization}

The efficiency of TP matching depends a lot on the size of the pattern \cite{AmerYahiaCLS01,Ramanan02,FlescaFM03,ChenC08}. It is thus essential to identify and eliminate redundant nodes in the pattern and do so as efficiently as possible~\cite{AmerYahiaCLS01}. This process is called TP minimization.

All research related to TP minimization is based on a pioneer paper by Amer-Yahia \emph{et al.} \cite{AmerYahiaCLS01}, who formulate the problem as follows: given a TP, find an equivalent TP of the smallest size. Formally, given a data tree $t$ and a TP $p$ of size (i.e., number of nodes) $n$, let $S = \{p_i\}$ be the set of TPs of size $n_i$ contained in $p$ ($p_i \subseteq p$ and $n_i \leq n~\forall i$). Minimizing $p$ is finding a TP $p_{min} \in S$ of size $n_{min}$ such that:
\begin{itemize}
	\item $p_{min} \equiv p$ when matched against $t$;
	\item $n_{min} \leq n_i~\forall i$.
\end{itemize}

Moreover, a set $C$ of integrity constraints (ICs) may be integrated into the problem. There are two forms of ICs:
\begin{enumerate}
	\item each node of type $A$ (e.g., book) must have a child (respectively, descendant) of type $B$ (e.g., author), denoted $A \to B$ (respectively, $A \Rightarrow B$);
	\item each node of type $A$ (e.g., book) must have a descendant of type $C$ (e.g., name), knowing that $C$ is a descendant of $B$ (e.g, author), i.e., $A \Rightarrow C$ knowing that $B \Rightarrow C$.
\end{enumerate}

Since TP minimization relies on the concepts of containment and equivalence, we first detail how containment and equivalence are tested (Section~\ref{sec:EquivalenceAndContainmentTests}). Then, we review the approaches that address the TP minimization problem without taking ICs into account (Section~\ref{sec:UnconstrainedMinimization}), and the approaches that do (Section~\ref{sec:MimizationUnderIntegrityConstraints}).

\begin{figure*}[hbt]
\centering
\includegraphics[width=13cm]{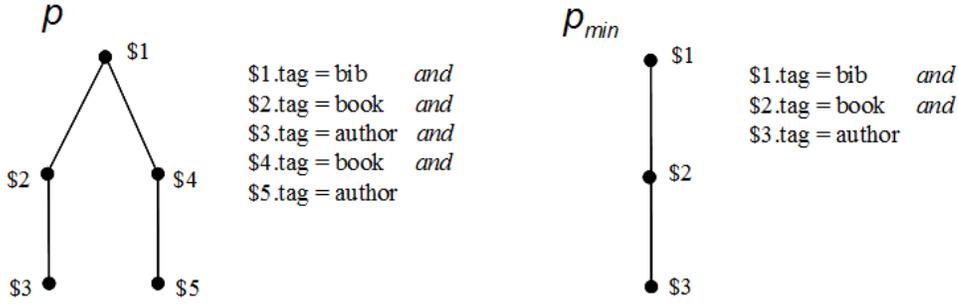} 
\caption{Simple minimization example}
\label{fig-minimization}
\end{figure*}

\subsubsection{Containment and equivalence testing}
\label{sec:EquivalenceAndContainmentTests}

Let us first further formalize the definitions of containment and equivalence. \emph{Containment} of two TPs $p_1$ and $p_2$ is defined as a node mapping relationship $h: p_1 \rightarrow p_2$ such that \cite{AmerYahiaCLS01,MiklauS02}:
\begin{itemize}
	\item $h$ preserves node type, i.e., $\forall x \in p_1$, $x$ and $h(x)$ must be of the same type. Moreover, if $x$ is an output node, $h(x)$ must be an output node too;
	\item $h$ preserves node relationships, i.e., if two nodes $(x, y)$ are linked through a PC (respectively, AD) relationship in $p_1$, $(f(x), f(y))$ must also be linked through a PC (respectively, AD) relationship in $p_2$.
\end{itemize}

Note that function $h$ is very similar to function $f$ that matches a TP to a data tree (Section~\ref{matching}), but here, $h$ is a homomorphism between two TPs \cite{ChandraM77}. Moreover, containment may be tested between a TP fragment and this TP as a whole. $h$ is then an endomorphism.

Finally, \emph{equivalence} between two TPs is simply considered as two-way containment \cite{AmerYahiaCLS01,FlescaFM03}. Formally, let $p_1$ and $p_2$ be two TPs. $p_1 \equiv p_2$ if and only if $p_1 \subseteq p_2$ and $p_2 \subseteq p_1$.

Miklau and Suciu show that containment testing is intractable in the general case \cite{MiklauS02}. They nonetheless propose an efficient algorithm for significant particular cases, namely TPs with AD edges and an output node. This algorithm is based on tree automata, the objective being to reduce the TP containment problem to that of regular tree languages. In summary, given two TPs $p_1$ and $p_2$, to test whether $p_1 \subseteq	p_2$:
\begin{itemize} 
	\item $p_1$'s nodes and edges are matched to deterministic finite tree automaton $A_1$'s states and transitions, respectively;
	\item $p_2$'s nodes and edges are matched to alternating finite tree automaton $A_2$'s states and transitions, respectively;
	\item if $lang(A_1) \subseteq lang(A_2)$, then $p_1 \subseteq	p_2$, where $lang(A_i)$ is the language associated with automaton $A_i~(i = \{1, 2\})$.
\end{itemize}

Similar approaches further test TP containment under Document Type Definition (DTD) constraints \cite{NevenS03,Wood03}. For instance, Wood exploits regular tree grammars (RTGs) to achieve containment testing \cite{Wood03}. Let $D$ be a DTD and $G_1$ and $G_2$ RTGs corresponding to $p_1$ and $p_2$, respectively. Then $p_1 \subseteq	p_2$ if and only if $(D \cap G_1) \subseteq (D \cap G_2)$.

\subsubsection{Unconstrained minimization}
\label{sec:UnconstrainedMinimization}

The first TP minimization algorithm, Constraint Independent Minimization (CIM) \cite{AmerYahiaCLS01}, eliminates redundant nodes from TP $p$ by exploiting the concept of images. Let there be a node $x \in p$. Its list of images, denoted $images(x)$, is composed of nodes from $p$ that bear the same type as $x$, but are different from $x$. For each leaf $x \in p$, CIM searches for nodes of the same type as $x$ in the set of descendants of $images(parent(x))$, where $parent(x)$ is $x$'s parent node in $p$. If such nodes are found, $x$ is redundant and thus deleted from $p$. CIM then proceeds similarly, in a bottom-up fashion, on $parent(x)$, until all nodes in $p$ (except output nodes that must always be retained) have been checked for redundancy. 

For example, let us consider TP $p$ from Figure~\ref{fig-minimization}. Without regarding node order, let us check leaf node \$5 for redundancy. $parent(\$5) = \$4, images(\$4) = \{\$2\}, descendants(\$2) = \{\$3\}$, where $descendants(x)$ is the set of descendants of node $x$. \$3 bears the same type as \$5 (author), thus \$5 is deleted. Similarly, \$4 is then found redundant and deleted, outputting $p_{min}$ in Figure~\ref{fig-minimization}. Further testing \$1, \$2 and \$3 does not detect any new redundant node. Moreover, deleting another node from $p_{min}$ would break the equivalence between $p$ and $p_{min}$. Hence, $p_{min}$ is indeed minimal.


All minimization algorithms subsequent to CIM retain its principle while optimizing complexity. One strategy is to replace images by more efficient relationships, i.e., coverage \cite{ChenC06,CheL08}, also called simulation \cite{Ramanan02}. Let $p$ be the TP to minimize, and $x, y \in p$ two of its nodes. If $y$ covers $x$ ($x \angle y$) \cite{AbiteboulV00}:
\begin{itemize}
	\item the types of $x$ and $y$ must be identical;
	\item if $x$ has a child (respectively descendant) node $x'$, $y$ must have a child (respectively descendant) node $y'$ such that $x' \angle y'$.
\end{itemize}
$cov(x)$ denotes the set of nodes that cover $x$. $cov(x) = x$ if $x$ is an output node. Then, the minimization process tests node redundancy in a top-down fashion as follows: $\forall x \in p$, $\forall x' \in children(x)$ (respectively $descendants(x)$), if $\exists x'' \in children(x)$ (respectively $descendants(x)$) such that $x'' \in cov(x')$, then $x'$ \emph{and the subtree rooted in $x'$} are deleted. $children(x)$ (respectively $descendants(x)$) is the set of direct children (respectively, all descendants) of node $x$.

Another strategy is to simply prune subtrees recursively \cite{ChanFFGR02}. The subtree rooted at node $x$, denoted $subtree(x)$, is minimized in two steps:
\begin{enumerate}
	\item $\forall x', x'' \in children(x)$, if $subtree(x') \subseteq subtree(x'')$ then delete $subtree(x')$;
	\item $\forall x' \in children(x)$ (remaining children of $x$), minimize $subtree(x')$.
\end{enumerate}
A variant proceeds similarly, by first searching in a TP $p$ for any subtree $p_i$ redundant with $sp$, where $sp$ is $p$ stripped of its root \cite{FlescaFM03}. Formally, the algorithm tests whether $p - sp_i \subseteq p_i$. Redundant subtrees are removed. Then, the algorithm is recursively executed on unpruned subtrees $sp_i$.

\subsubsection{Minimization under integrity constraints}
\label{sec:MimizationUnderIntegrityConstraints}

Taking ICs into account in the minimization process is casually achieved as follows.
\begin{enumerate}
	\item Augment the TP to minimize with nodes and edges that represent ICs. This is casually achieved with the classical chase technique \cite{Ullman88}. For instance, if we had a book $\to$ author IC (nodes of type ``book'' must have a child of type ``author''), we would add one child node of type author to each of the nodes \$2 and \$4 from Figure~\ref{fig-minimization}. Note that augmentation must only apply to nodes from the original TP, and not to nodes previously added by the chase.
	\item Run any TP minimization algorithm, without testing utilitarian nodes introduced in step~\#1 for redundancy, so that ICs hold.
	\item Delete utilitarian nodes introduced in step \#1.
\end{enumerate}
This process has been applied onto the CIM algorithm, to produce ACIM (Augmented CIM) \cite{AmerYahiaCLS01}, as well as on its coverage-based variants \cite{Ramanan02,ChenC06,ChenC06a}. Since the size of an augmented TP can be much larger than that of the original TP, an algorithm called Constraint Dependant Minimization (CDM) also helps identify and prune all TP nodes that are redundant under ICs \cite{AmerYahiaCLS01}. CDM actually acts as a filter before ACIM is applied. CDM considers a TP leaf $x'$ of type $T'$ redundant and removes it if one of the following conditions holds.
\begin{itemize}
	\item $parent(x') = x$ (respectively $ancestor(x') = x$) of type $T$ and there exists an IC $T \to T'$ (respectively $T \Rightarrow T'$).
	\item $parent(x') = x$ (respectively $ancestor(x') = x$) of type $T$, $\exists x'' / parent(x'') = x$ (respectively $ancestor(x'') = x$) of type $T''$, and there exists an IC $T' = T''$ (respectively, there exists one of the ICs $T'' \Rightarrow T'$ or $T' = T''$).
\end{itemize}
An alternative simulation-based minimization algorithm also includes a similar prefilter, along with an augmentation phase that adds to the nodes of $cov(x)$ their ancestors instead of using the chase \cite{Ramanan02}. 

Finally, the scope of ICs has recently been extended to include not only forward and subtype (FT) constraints (as defined in Section~\ref{sec:PatternTreeMinimization}), but also backward and sibling (BS) constraints \cite{ChenC08}:
\begin{itemize}
	\item each node of type $A$ (e.g., author) must have a parent (respectively, ancestor) of type $B$ (e.g., book), i.e., $A \gets B$ (respectively, $A \Leftarrow B$);
	\item each node of type $A$ (e.g., book) that has a child of type $B$ (e.g., editor) must also have a child of type $C$ (e.g, address), i.e., if $A \to B$ then $A \to C$.
\end{itemize}
Under FBST constraints, several minimal TPs can be achieved (\emph{vs.} one only under FT constraints), which allows further optimizations of the chase augmentation and simulation-based minimization processes.

\subsection{Holistic tree pattern matching}
\label{sec:holistic}

While TP minimization approaches (Section~\ref{sec:PatternTreeMinimization}) wholly focus on the TP side of the matching process, holistic matching (also called holistic twig join \cite{GrimsmoB10}) algorithms mainly operate on minimizing access to the input data tree when performing actual matching operations. The initial binary join-based approach for matching proposed by Al-Khalifa \emph{et al.} \cite{Al-KhalifaJPWKS02} indeed produces large intermediate results. Holistic approaches casually optimize TP matching in two steps \cite{Lu10Dasfaa}:
\begin{enumerate}
	\item \emph{labeling:} assign to each node $x$ in the data tree $t$ an integer label $label(x)$ that captures the structure of $t$ (Section~\ref{sec:LabelingPhase});
	\item \emph{computing:} exploit labels to match a twig pattern $p$ against $t$ without traversing $t$ again (Section~\ref{sec:ComputingPhase}).
\end{enumerate}
Moreover, recent proposals aim at reducing data tree size by exploiting structural summaries in combination to labeling schemes. We review them in Section~\ref{sec:StructuralSummaryBasedApproaches}.

Let us finally highlight that, given the tremendous number of holistic matching algorithms proposed in the literature, it is quite impossible to review them all. Hence, we aim in the following sections at presenting the most influential. The interested reader may further refer to Grimsmo and Bj{\o}rklund's survey \cite{GrimsmoB10}, which uniquely focuses on and introduces a nice history of holistic approaches.

\subsubsection{Labeling phase}
\label{sec:LabelingPhase}

The aim of data tree labeling schemes is to determine the relationship (i.e., PC or AD) between two nodes of a tree from their labels alone \cite{Lu10Dasfaa}. Many labeling schemes have been proposed in the literature. We particularly focus in this section on the region encoding (or containment) and the Dewey ID (or prefix) labeling schemes that are used in holistic approaches. However, other approaches do exist, based on a tree-traversal order \cite{LiM01}, prime numbers \cite{WuLH04} or a combination of structural index and inverted lists \cite{KaushikKNR04}, for instance.

\begin{figure*}[hbt]
\centering
\includegraphics[width=16cm]{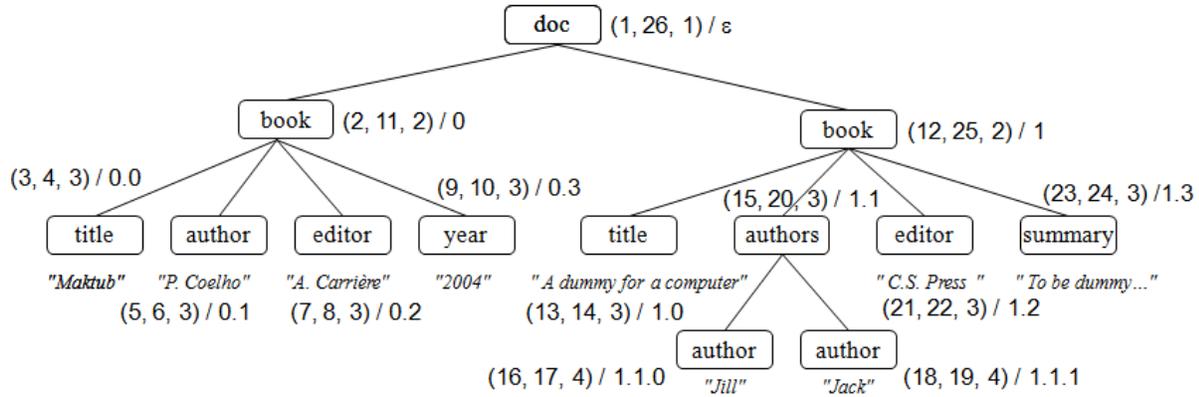} 
\caption{Sample data tree labeling}
\label{fig-labeling}
\end{figure*}

The region encoding scheme \cite{BrunoKS02} labels each node $x$ in a data tree $t$ with a 3-tuple $(start, end,$ $level)$, where $start$ (respectively, $end$) is a counter from the root of $t$ until the start (respectively, the end) of $x$ (in depth first), and $level$ is the depth of $x$ in $t$. For example, the data tree from Figure~\ref{fig-runex}(a) is labeled in Figure~\ref{fig-labeling}, with the region encoding scheme label indicated between parentheses. In Figure~\ref{fig-labeling}, node $x (title =$ ``Maktub''$)$ is labeled (3, 4, 3). $start = 3$ because $x$ is the first child of the node with $start = 2$; $end = 4$ because $x$ has no children (thus $end = start + 1$); and $level = 3$ (level 1 being the root's).

Now, let $x$ and $x'$ be two nodes labeled $(S, E, L)$ and $(S', E', L')$, respectively. Then:
\begin{itemize}
	\item $x'$ is a descendant of $x$ if and only if $S < S'$ and $E' < E$;
	\item $x'$ is a child of $x$ if and only if $S < S'$, $E' < E$ and $L' = L + 1$.
\end{itemize}
For example, in Figure~\ref{fig-labeling}, node $(author =$ ``Jack''$)$ labeled (18, 19, 4) is a descendant from the node $book$ labeled (12, 25, 2) and a child of node $authors$ labeled (15, 20, 3).

The Dewey ID scheme \cite{TatarinovVBSSZ02} labels tree nodes as a sequence. The root node is labeled $\epsilon$ (empty). Its children are labeled 0, 1, 2, etc. Then, at any subsequent level, the children of node $x$ are labeled $label(x).0$, $label(x).1$, $label(x).2$, etc. More formally, for each non-root element $x'$, $label(x') = label(x).i$, where $x'$ is the $i^{th}$ child of $x$. Thus, each label embeds all ancestors of the associated node. For example, in Figure~\ref{fig-labeling}, the Dewey ID label is featured on the right hand side of the ``/'', for each node. Node $(author =$ ``Jack''$)$, labeled 1.1.1, is the second child of the node labeled 1.1 (i.e., $authors$), and a descendant of the node labeled 1 (i.e., the right hand side $book$).

The Dewey ID scheme has been extended to incorporate node names \cite{LuLCC05}, by exploiting schema information available in a DTD or XML schema. Encoding node names along a path into a Dewey label provides not only the labels of the ancestors of a given node, but also their names. Moreover, the Dewey ID scheme suffers from a high relabeling cost for dynamic XML documents where nodes can be arbitrarily inserted and deleted. Thus, variant schemes, namely ORDPATH \cite{ONeilOPCSW04} and Dynamic DEwey (DDE) \cite{XuLWB09}, have been devised to dynamically extend the domain of label component values, so that no global relabeling is required.

The main difference between the region encoding and Dewey ID labeling schemes lies in the way structural relationships can be inferred from a label. While region encoding necessitates two nodes to determine whether they are related by a PC or AD relationship, Dewey IDs directly relate to ancestors and thus only require to know the current node's label. A Dewey ID-labeled data tree is also easier to update than a region encoded data tree \cite{Lu06thesis}.

\subsubsection{Computing phase}
\label{sec:ComputingPhase}

Various holistic algorithms actually achieve TP matching, but they all exploit a data list that, for each node, contains all labels of nodes of the same type. In this section, we first review the approaches based on the region encoding scheme, which were first proposed, and then the approaches based on the Dewey ID scheme.

\begin{figure}[hbt]
\centering
\includegraphics[width=8.45cm]{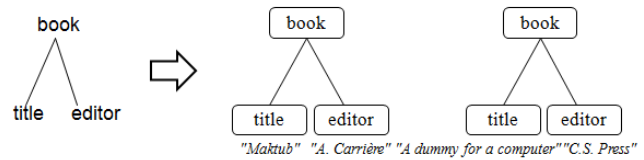}
\caption{Sample holistic matching}
\label{fig-holistic}
\end{figure}

As their successors, the first popular holistic matching algorithms, PathStack and TwigStack, proceed in two steps \cite{BrunoKS02}: intermediate path solutions are output to match each query root-to-leaf path, and then merged to obtain the final result. For example, let us consider the TP represented on the left hand side of Figure~\ref{fig-holistic}, to be matched against the data tree from Figure~\ref{fig-labeling}. Intermediate path solutions follow, expressed as labels.
\begin{itemize}
	\item $book/title$: (2, 11, 2) (3, 4, 3), (12, 25, 2) (13, 14, 3)
	\item $book/editor$: (2, 11, 2) (7, 8, 3), (12, 25, 2) (21, 22, 3)
\end{itemize}

After merging these intermediate paths, we obtain the label paths below, which correspond to the witness trees represented on the right hand side of Figure~\ref{fig-holistic}.
\begin{itemize}
	\item (2, 11, 2) (3, 4, 3) (7, 8, 3)
	\item (12, 25, 2) (13, 14, 3) (21, 22, 3)
\end{itemize}

One issue with TwigStack is that it only considers AD relationships in the TP and does not consider level information. Hence, it may output many useless intermediate results for queries with PC relationships. Moreover, it cannot process queries with order predicates. 

On one hand, in order to reduce the search space, path summaries that exploit schema information may be used \cite{BartaCM05}. If a pattern subtree matches a data tree several times, TwigStack loads streams for all distinct paths, whether these streams contribute to the output or not. Path summaries help distinguish whether the occurrences of each pattern subtree are output elements or not. Those that do not are pruned.

On the other hand, TwigStackList better controls the size of intermediate results by buffering parent elements in PC relationships in a main-memory list structure \cite{LuCL04}. Thus, only AD relationships in branching edges\footnote{A \emph{branching node} is a node whose number of children is greater than one. All edges originating from a branching node are called \emph{branching edges} \cite{LuCL04}.} are handled by TwigStackList, and not in all edges as with TwigStack. OrderedTJ builds upon TwigStackList by handling order specifications in TPs \cite{LuLYLN05}. OrderedTJ additionally checks the order conditions of nodes before outputting intermediate paths, with the help of a stack data structure.

Since TwigStack and OrderedTJ partition data to streams according to their names alone, two new data streaming techniques are introduced in iTwigJoin \cite{ChenLL05}: the tag+level and prefix path schemes. In the OrderedTJ algorithm, only AD relationships in branching edges are taken into account. The tag+level scheme also takes PC relationships in all edges into account. The prefix path scheme further takes 1-branching into account. 

An eventual enhancement has been brought by Twig$^{2}$Stack, which optimizes TwigStack by further reducing the size of intermediate results \cite{ChenLTHAC06}. Twig$^{2}$Stack associates each query node $x$ with a hierarchical stack. A node $x'$ is pushed into hierarchical stack $HS[x]$ if and only if $x'$ satisfies the sub-twig query rooted at $x$. Matching can be determined when an entire sub-tree of $x'$ is seen with respect to post-order data tree traversal. Baca \emph{et al.} also fuse TwigStack and Twig$^{2}$Stack along the same line, still to reduce the size of intermediate results \cite{BacaKS08}.

Simply replacing the region encoding labeling scheme by the Dewey ID scheme would not particularly improve holistic matching approaches, since they would also need to read labels for all tree nodes. However, exploiting the \emph{extended} Dewey labeling scheme allows further improvements.

TJFast constructs, for each node $x$ in the TP, an input stream $T_x$ \cite{LuLCC05}. $T_x$ contains the ordered extended Dewey labels of nodes of the same type as $x$. As TwigStackList, TJFast assigns, for each branching node $b$, a set of nodes $S_b$ that are potentially query answers. But with TJFast, the size $S_b$ is always bounded by the depth of the data tree. TJFast+L further extends TJFast by including the tag+level streaming scheme \cite{Lu06thesis}.

Eventually, Lu \emph{et al.} have recently identified a key issue in holistic algorithms, called matching cross \cite{LuLBCW-TKDE10}. If a matching cross is encountered, a holistic algorithm either outputs useless intermediate results and becomes suboptimal, or misses useful results and looses its matching power. Based on this observation, the authors designed a set of algorithms, collectively called TreeMatch, which use a concise encoding to present matching results, thus reducing useless intermediate results. TreeMatch has been proved optimal for several classes of queries based on extended TPs (Section~\ref{extp}).

\subsubsection{Structural summary-based approaches}
\label{sec:StructuralSummaryBasedApproaches}

\begin{figure}[hbt]
\centering
\includegraphics[width=8.45cm]{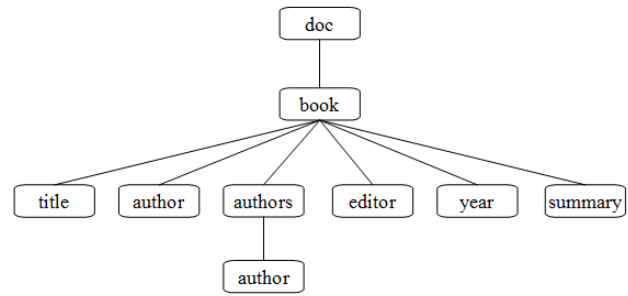}
\caption{Sample DataGuide}
\label{fig-dataguide}
\end{figure}

These approaches aim at avoiding repeated access to the input data tree. Thus, they exploit structural summaries similar to the DataGuide proposed for semi-structured documents \cite{GoldmanW97}. A DataGuide's structure describes using one single label all the nodes whose labels (types) are identical. Its definition is based on targeted path sets, i.e., sets of nodes that are reached by traversing a given path. For example, the DataGuide corresponding to the data tree from Figure~\ref{fig-runex}(a) is represented in Figure~\ref{fig-dataguide}.

While a DataGuide can efficiently answer queries with PC edges (by matching the query path against the label path directly), it cannot process queries with AD edges nor twig queries, because it does not preserve hierarchical relationships \cite{HawL09}. Combining a DataGuide with a labeling scheme that captures AD relationships allows the reconstruction, for any node $x$ in a data tree $t$, of the specific path instance $x$ belongs to. Thus, in the computing phase, node labels can be compared without accessing labels related to inner TP nodes \cite{IzadiHH09}.

For instance, TwigX-Guide combines a DataGuide to the region encoding labeling scheme \cite{HawL09}. Version Tree is an annotated DataGuide in which labels include a version number for nodes of the same type (e.g., all $book$ nodes) \cite{WuL08}. Combined to the Dewey ID labeling scheme, it supports a matching algorithm called TwigVersion. Finally, QueryGuide labels DataGuide nodes with Dewey ID lists and is part of the S$^3$ matching method \cite{IzadiHH09}. All three approaches have been experimentally shown to perform matching faster than previous holistic algorithm such as TwigStack and TJFast.

\subsection{Other pattern tree matching approaches}
\label{sec:SpecificApproaches}

We present in this section matching approaches of interest that do not fall into the minimization and holistic families of methods, namely tree homeomorphism matching (Section~\ref{sec:TreeHomeomorphismMatching}) and TP relaxation (Section~\ref{sec:TreePatternRelaxation}).

\subsubsection{Tree homeomorphism matching}
\label{sec:TreeHomeomorphismMatching}

The tree homeomorphism matching problem is a particular case of the TP matching problem. More precisely, the considered TPs only bear descendant edges. Formally, given a TP $p$ and a data tree $t$, tree homeomorphism matching aims at determining whether there is a mapping $\theta$ from the nodes of $p$ to the nodes $t$ such that if node $x'$ is a \emph{child} of $x$ in $p$, then $\theta(x')$ is a \emph{descendant} of $\theta(x)$ in $t$.

G{\"o}tz \emph{et al.} propose a series of algorithms that aim at reducing the time and space complexity of previous homeomorphism matching algorithms \cite{GotzKM09}. Their whole work is based on a simple matching procedure called {\sc Match}. Let $x$ be a node of TP $p$ and $y$ a node of data tree $t$. {\sc Match} tests whether the subtree rooted at $x$, $subtree(x)$, matches $subtree(y)$. If $y$ matches $x$, children of $x$ are recursively tested to match any child of $y$. If $y$ does not match $x$, then $x$ is recursively tested to match any child of $y$. {\sc Match} uses the recursion stack to determine which function call to issue next or which final value to return, e.g., to determine the data node $y$ onto which $x$'s parent was matched in $t$ before proceeding with $x$'s siblings. In opposition, {\sc L-Match} recomputes the information necessary to make the decision with a backtracking function. 

{\sc Match} and {\sc L-Match} are space-efficient top-down algorithms, but involve a lot of recomputing and thus bear a high time complexity. Thus, G{\"o}tz \emph{et al.} also introduce a bottom-up strategy. It is based on algorithm {\sc TMatch} that addresses the tree homeomorphism problem. {\sc TMatch} exploits a left-to-right post-order ordering $<_{post}$ on nodes, and returns the largest (w.r.t. $<_{post}$) TP node $x$ in an interval $[x_{from}, x_{until}]$ (still w.r.t. $<_{post}$) such that $subtree(y)$ matches $[x_{from}, x]$ if $x$ exists; and $x_{from} - 1$ (the predecessor of $x_{from}$ w.r.t. $<_{post}$) otherwise. Finally, {\sc TMatch-All} generalizes {\sc TMatch} to address the tree homeomorphism matching problem, i.e., {\sc TMatch-All} computes all possible exact matches of $p$ against $t$.

\subsubsection{Tree pattern relaxation}
\label{sec:TreePatternRelaxation}

\begin{figure}[hbt]
\centering
\includegraphics[width=8.45cm]{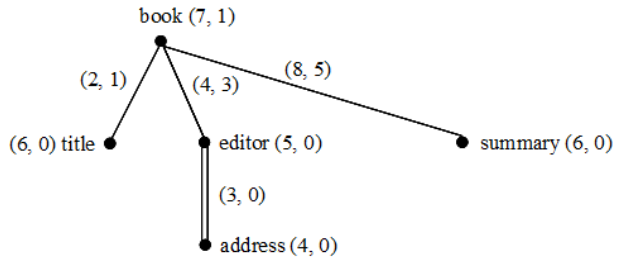}
\caption{Sample weighted tree pattern}
\label{fig-weighted}
\end{figure}

TP relaxation is not an optimization of the matching process \emph{per se}, but an optimization of its result with respect to user expectations. TP relaxation indeed allows approximate TP matching and returns ranked answers in the spirit of Information Retrieval \cite{AmerYahiaCS02}. Four TP relaxations are proposed, the first two relating to structure and the last two to content.
\begin{enumerate}
	\item \emph{Edge generalization} permits a PC edge in the TP to be generalized to an AD edge. For example, in Figure~\ref{fig-weighted}, the $book/title$ edge can be generalized to allow books with any descendant $title$ node to appear in the witness tree.
	\item \emph{Subtree promotion} permits to connect a whole subtree to its grandparent by an AD edge. For example, in Figure~\ref{fig-weighted}, the $address$ node can be promoted to allow $book$ nodes that have a descendant $address$ node to be output even if the $address$ node is not a descendant of the $editor$ node.
	\item \emph{Leaf node deletion} permits a leaf node to be deleted. For example, in Figure~\ref{fig-weighted}, the $summary$ node can be deleted, allowing for books to appear in the witness tree whether they bear a summary or not.
	\item \emph{Node generalization} permits to generalize the type of a query node to a supertype. For example, in Figure~\ref{fig-weighted}, the node $book$ could be generalized to node $doc$ (document) from Figure~\ref{fig-runex}(a).
\end{enumerate}

Answer ranking is achieved by computing a score. To this aim, the weighted TP is introduced, where each node and edge is associated with an exact weight $ew$ and a relaxed weight $rw$ such that $ew \geq rw$.  Figure~\ref{fig-weighted} features a sample weighted TP. In this example, the score of exact matches of the weighted TP is equal to the sum of the exact weight of its nodes and edges, i.e., 41. If the node $book$ was generalized to $doc$, the score of an approximate answer that is a document is the sum of $rw(book)$ and the exact weight of the other nodes and edges, i.e., 35.

\subsection{Discussion}
\label{sec:Discussion2}

In this section, we discuss and compare the TP matching optimization approaches surveyed in Sections~\ref{sec:PatternTreeMinimization}, \ref{sec:holistic} and \ref{sec:SpecificApproaches}; except TP relaxation, whose goal is quite different (i.e., augmenting user satisfaction rather than matching efficiency). Choosing objective comparison criteria is pretty straightforward for algorithms. \emph{Time} and \emph{space complexity} immediately come to mind, though they are diversely documented in the papers we survey. 

We could also retain algorithm \emph{correctness} as a comparison criterion, although proofs are quite systematically provided by authors of matching optimization approaches. Thus, we only notice here that none of the minimization algorithms reviewed in Section~\ref{sec:PatternTreeMinimization} actually test the equivalence of minimal TP $p_{min}$ to original TP $p$. More precisely, containment (i.e., $p_{min} \subseteq p$) is checked \emph{a posteriori}, but the minimization process being assumed correct, equivalence is not double-checked.

Algorithm comparison with respect to complexity follows (Sections~\ref{sec:TimeComplexity} and \ref{sec:SpaceComplexity}), and is synthesized in Section~\ref{sec:Synthesis2}, where we also further discuss the complementarity between TP minimization and holistic approaches.

\subsubsection{Time complexity}
\label{sec:TimeComplexity}

Time complexity is quite well documented for minimization approaches. Except the first, naive matching algorithms \cite{AmerYahiaCLS01}, all optimized minimization algorithms, whether they take ICs into account or not, have a worst-case time complexity of $O(n^2)$, where $n$ is the size (number of nodes) of the TP. A notable exception is Chen and Chan's extension of ICs to FSBT constraints~\cite{ChenC08} that makes the matching problem more complex. Thus, their algorithms range in complexity, with respect to the combination of ICs that are processed, from $O(n^2)$ (F and FS ICs) to $O(n^3 \cdot s + n^2 \cdot s^2)$ (FSBT ICs), where $s$ is the size of the set of element types in ICs.

On the other hand, in papers about holistic approaches, complexity is often disregarded in favor of the class of TP a given algorithm is optimal for. The class of TP may be \cite{Baca08thesis}:
\begin{itemize}
	\item \emph{AD:} a TP containing only AD edges, which may begin with a PC edge;
	\item \emph{PC:} a TP containing only PC edges, which may begin with an AD edge;
	\item \emph{one-branching node TP:} a TP containing at most one branching node;
	\item \emph{AD in branches TP:} a TP containing only AD edges after the first branching node;
	\item \emph{PC before AD TP:} a TP where an AD edge is never found before a PC edge.
\end{itemize}

Holistic approaches indeed all have the same worst-case time complexity. It is linear in the sum of all sizes of the input lists they process \emph{and} the output list. However, since these approaches match twigs, they enumerate all root-to-leaf path matches. Thus, their complexity is actually exponential in $n$, i.e., it is $O(d^n)$, where $d$ is the size (number of nodes) of the input data tree \cite{ChenLTHAC06}. Only Twig$^2$Stack has a lower complexity, in $O(d \cdot b)$, where $b = max(b_1, b_2)$ with $b_1$ being the maximum number of query nodes with the same label and $b_2$ the maximum total number of children of query nodes with the same labels ($b \leq n$) \cite{ChenLTHAC06}; as well as TreeMatch \cite{LuLBCW-TKDE10}, which has been experimentally shown to outperform Twig$^2$Stack.

\begin{table*}[hbt]
\renewcommand{\arraystretch}{1.3}
\caption{Synthesis of tree pattern matching approaches}
\label{tab-synth-matching}
\centering
\begin{tabular}{|l|c|c|c|c|}
\hline
\textbf{Algorithm} & \textbf{TP used} & \textbf{Time complexity} & \textbf{Space complexity} & \textbf{Features} \\
\hline \hline
\textbf{CIM} \cite{AmerYahiaCLS01} & With output node & $O(n^4)$ & N/A & Prunes nodes  \\
\hline
\textbf{Coverage} \cite{CheL08} & With output node & $O(n^2)$ & N/A & Prune subtrees  \\ 
\textbf{Simulation} \cite{Ramanan02} & & & & \\
\hline
\textbf{Recursive pruning} \cite{ChanFFGR02} & Unspecified & $O(n^2)$ & N/A & Prune subtrees \\ 
\cite{FlescaFM03} & Limited branched TP & & & \\
\hline
\textbf{ACIM} \cite{AmerYahiaCLS01} & With output node & $O(n^6)$ & N/A & FS ICs \\
\hline
\textbf{CDM} \cite{AmerYahiaCLS01} & With output node & $O(n^2)$ & N/A & FS ICs \\
\hline
\textbf{Chase + coverage} \cite{CheL08} & With output node & $O(n^2)$ & N/A & FS ICs \\
\textbf{Chase + simulation} \cite{Ramanan02} & & & & \\
\hline
\textbf{FSBT} \cite{ChenC08} & With output node & From $O(n^2)$ to & N/A & FSBT ICs  \\ 
& & $O(n^3 \cdot s + n^2 \cdot s^2)$ & &  \\
\hline \hline
\textbf{TwigStack} \cite{BrunoKS02} & Twig pattern & $O(d^n)$ & $min(n \cdot l_t, s)$ & Region enc. \\
\hline
\textbf{TwigStackList} \cite{LuCL04} & Twig pattern & $O(d^n)$ & $O(n \cdot l_t)$ & Region enc. \\
\textbf{OrderedTJ} \cite{LuLYLN05} & Ordered twig &  & & \\
\textbf{iTwigJoin} \cite{ChenLL05} & Twig pattern &  & & \emph{Tag+level}, \emph{Prefix} \\
\hline
\textbf{Twig$^{2}$Stack} \cite{ChenLTHAC06} & GTP & $O(d \cdot b)$ & $O(d \cdot b)$ & Region enc. \\
\hline
\textbf{TJFast} \cite{LuLCC05} & Twig pattern & $O(d^n)$ & $O(l_t^2 \cdot bf + l_t \cdot f)$ & Dewey ID \\
\hline
\textbf{TreeMatch} \cite{LuLBCW-TKDE10} & Extended TP & $O(d \cdot b)$ & $O(l_t^2 \cdot bf + l_t \cdot f)$ & Dewey ID \\
\hline
\textbf{TwigX-Guide} \cite{HawL09} & Twig pattern & Unspecified & Unspecified & DataGuide + \\
& & & & Region enc. \\
\hline
\textbf{S$^3$} \cite{IzadiHH09} & TP with & Unspecified & Unspecified & DataGuide + \\
& logical operators & & & Dewey ID \\ 
\hline
\textbf{TwigVersion} \cite{WuL08} & Twig with & $O(l_v \cdot n_v + l_p \cdot e)$ & $O(l_v \cdot n_v)$ & DataGuide + \\
& node predicates & & & Dewey+Version \\
\hline \hline
\textbf{Tree homeomorphism} \cite{GotzKM09} & Unranked TP & $O(n \cdot d \cdot l_p)$ & $O(l_t \cdot log(bf))$ & Left-to-right \\
& & & & post-order \\
\hline 
\end{tabular}
\end{table*}

Structural summary-based approaches (Section~\ref{sec:StructuralSummaryBasedApproaches}) have been experimentally shown to perform matching faster than previous holistic algorithms such as TwigStack and TJFast. However, they have neither been compared to one another, nor to TreeMatch. Moreover, their time complexity is expressed in terms of DataGuide degree and specific features such as the number of versions in TwigVersion \cite{WuL08}, so it is not easy to directly compare it to other holistic algorithms'. For instance, TwigVersion has a worst case time complexity of $O(l_v \cdot n_v + l_p \cdot e)$, where $l_v$ is the depth of version tree (annotated DataGuide) $v$, $n_v$ is the total number of versions on nodes of $v$ whose tags appear in the leaf nodes of TP $p$, $l_p$ the depth of TP $p$, and $e$ is the size of the edge set containing all edges that are on paths from nodes of $v$ whose tags appear in the leaf nodes of $p$ to the root node of $v$ \cite{WuL08}.

Finally, the complexity of tree homeomorphism matching is proved to be $O(n \cdot d \cdot l_p)$ \cite{GotzKM09}.

\subsubsection{Space complexity}
\label{sec:SpaceComplexity}

Space complexity is intensively addressed in the literature regarding holistic approaches, which can produce many intermediate results whose volume must be minimized, so that algorithms can run in memory with as low response time as possible. On the other hand, space complexity is not considered an issue in minimization processes, which prune nodes in PTs that are presumably small enough to fit into any main memory.

Regarding holistic approaches, the worst-case space complexity of TwigStack is $min(n \cdot l_t, s)$ \cite{BrunoKS02}, where $n$ is defined as in Section~\ref{sec:TimeComplexity}, $l_t$ is the depth of data tree $t$ and $s$ is the sum of sizes of the $n$ input lists in the computing phase (Section~\ref{sec:ComputingPhase}). Subsequent holistic algorithms are more time-efficient than Twigstack because they are more space-efficient. TwigStackList and iTwigJoin both have a worst-case space complexity of $O(n \cdot l_t)$ \cite{LuCL04,ChenLL05}. Noticeably, Twig$^2$Stack's space complexity is indeed the same as its time complexity, i.e., $O(d \cdot b)$ (Section~\ref{sec:TimeComplexity}). Finally, TJFast and TreeMatch have a worst-case space complexity of $O(l_t^2 \cdot bf + l_t \cdot f)$, where $bf$ is the maximal branching factor of the input data tree and $f$ the number of leaf nodes in the TP \cite{LuLCC05,LuLBCW-TKDE10}.

Structural summary approaches' space complexity, as is the case for time complexity, are difficult to compare because of their specifics. For example, the worst case space
complexity of TwigVersion is $O(l_v \cdot n_v)$, where $l_v$ and $n_v$ are defined as in Section~\ref{sec:TimeComplexity} \cite{WuL08}. Eventually, tree homeomorphism matching algorithms have a space complexity of $O(l_t \cdot log(bf))$ \cite{GotzKM09}.

\subsubsection{Synthesis}
\label{sec:Synthesis2}

Table~\ref{tab-synth-matching} recapitulates the characteristics of all matching optimization algorithms surveyed in this paper. In addition to the complexity comparison criteria discussed above, we also indicate in Table~\ref{tab-synth-matching} the type of TP handled by each algorithm, in reference to Section~\ref{sec:PatternTreeStructures}, as well as each algorithm's distinctive features. The first third of Table~\ref{tab-synth-matching} is dedicated to minimization approaches, the second to holistic approaches, and the third to tree homeomorphism matching.


In the light of this synthesis, we can notice again that TP minimization and holistic approaches have developed along separate roads. Papers related to one approach seldom refer to the other. Indeed, these two families of approaches cannot be compared, e.g., in terms of raw complexity, since TP minimization operates on TPs only, while holistic approaches optimize the actual matching of a TP against a data tree. TP minimization is actually implicitly considered as preprocessing TPs before matching \cite{YaoZ04}. We nonetheless find it surprising that nobody has ever combined TP minimization to holistic matching in a single framework to benefit from optimization of both data tree access and TP size. 

Eventually, Table~\ref{tab-synth-matching} clearly outlines the history and outcome of the families of algorithms we survey. With respect to PT minimization algorithms, unless elaborated ICs (i.e., FSBT ICs) are needed, the choice should clearly fall on an approach whose time complexity is $O(n^2)$, such as the coverage-based approaches \cite{CheL08,Ramanan02}. With respect to holistic matching, the raw efficiency of algorithms is not always easy to compare, especially between the latest descendants of Twigstack and structural summary-based approaches such as TwigVersion \cite{WuL08} or S$^3$ \cite{IzadiHH09}, for both complexity and experimental studies remain partial as of today. The types of TPs (Section~\ref{sec:TimeComplexity}) for which an approach is optimal thus remains a primary criterion. However, we agree with Grimsmo and Bj{\o}rklund in stating that a global holistic approach, i.e., an approach that encompasses all kinds of TPs, is most desirable \cite{GrimsmoB10}. In this respect, TreeMatch \cite{LuLBCW-TKDE10} appears as the most comprehensive solution as of today.

\begin{figure*}[hbt]
\centering
\includegraphics[width=15cm]{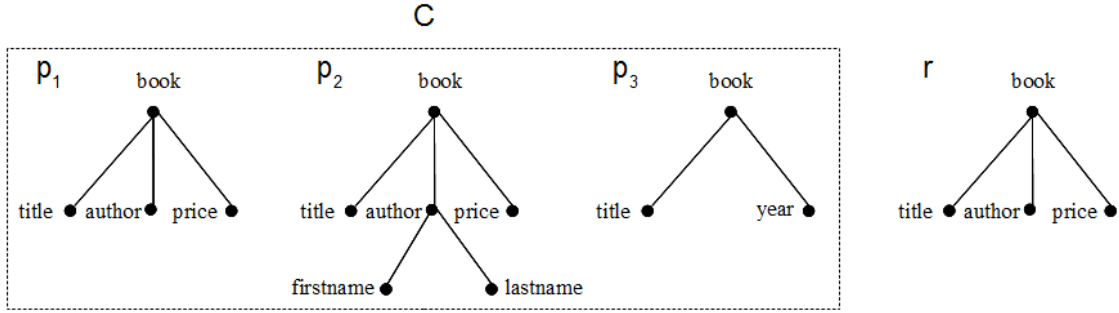} 
\caption{Sample tree pattern collection and rooted subtree}
\label{fig-mining}
\end{figure*}

\section{Tree pattern usages}
\label{sec:PatternTreeUsages}

\IEEEPARstart{B}{eside} expressing and optimizing queries over tree-structured documents, TPs have also been exploited for various purposes ranging from system optimization (e.g., query caching \cite{YangLH03vldb,WangWL09}, addressing and routing over a peer-to-peer network \cite{TatarinovH04}) to high-level database operations (e.g., schema construction \cite{ZhangLBT03},  active XML query satisfiability and relevance \cite{MaHZ08,AbiteboulBM09}) and knowledge discovery (e.g., discovering user communities \cite{ChandFG07}).

In this section, we investigate the most prominent of TP usages we found in the literature, which we classify by the means used to achieve the goals we have listed above (e.g., routing and query satisfiability), i.e., TP mining (Section~\ref{sec:tpmining}), TP rewriting (Section~\ref{sec:tpqreformulation}) and extensions to matching (Section~\ref{sec:xmatching}).

\subsection{Tree pattern mining}
\label{sec:tpmining}

TP mining actually summarizes into discovering frequent subtrees in a collection of TPs. It is used, for instance, to cache the results of frequent patterns, which significantly improves query response time \cite{YangLH03vldb}, produce data warehouse schemas of integrated XML documents from historical user queries \cite{ZhangLBT03}, or help in website management by mining data streams \cite{HsiehWC06}.

\subsubsection{Problem formulation}
\label{sec:ProblemFormulation}

Let $C = \{p_1, p_2, ..., p_n\}$ be a collection of $n$ TPs and $minsup \in~]0, 1]$ a number called minimum support. The support of any rooted subtree (denoted RST) $r$ is $sup(r) = freq(r) / n$, where $freq(r)$ is the total occurrence of $r$ in $C$. Then, the problem of mining frequent RSTs from $C$ may be defined as finding the set $F = \{r_1, r_2, ..., r_m\}$ of $m$ RSTs such that $\forall i \in [1, m], sup(r_i) \geq minsup$ \cite{YangLH03vldb}.

For example, let us consider TP collection $C$ and RST $r$ from Figure~\ref{fig-mining}. Since $r \subseteq p_1$ and $r \subseteq p_2$, $freq(r) = 2$ and $sup(r) = \frac{2}{3}$. If $minsup = \frac{1}{2}$, then $r$ is considered frequent. Note that searching frequent RSTs relies on testing containment (Section~\ref{sec:EquivalenceAndContainmentTests}) against TPs of $C$.

\subsubsection{Frequent subtree mining algorithms}
\label{sec:FrequentSubtreeMiningAlgorithms}

One of the first frequent RST mining algorithm, XQPMiner \cite{YangLHA03}, operates like the famous frequent itemset mining algorithm Apriori \cite{AgrawalS94}. XQPMiner initializes by enumerating all frequent 1-edge RSTs. Then, each further step $i$ is subdivided in two substeps: 
\begin{enumerate}
	\item candidate $i$-edge RSTs are built from $(i-1)$-edge RSTs and filtered with respect to the minimum support;
	\item each remaining $i$-edge RST is tested for containment in each TP of $C$ to compute its support.
\end{enumerate}

FastXMiner optimizes this process by constructing a global query pattern tree (G-QTP, Section~\ref{GQPT}) over $C$. Then, a tree-encoding scheme is applied on the G-QTP to partition candidate RSTs into equivalence classes. The authors show that only single-branch candidate RSTs need to be matched against the TPs from $C$, which leads to a large reduction in the number of tree 
inclusion tests \cite{YangLH03vldb}.

MineFreq further builds upon this principle by mining frequent RST \emph{sets} \cite{ZhangLBT03}. A frequent RST set must satisfy two requirements:  
\begin{enumerate}
	\item \emph{support requirement:} $sup(r_1, r_2, ..., r_n) \geq minsup$;
	\item \emph{confidence requirement:} $\forall r_i, \frac{freq(r_1, r_2, ..., r_n)}{freq(r_i)} \geq minconf$, where $minconf$ is a minimum confidence user-specified threshold.
\end{enumerate}
The algorithm again proceeds by level, the $n$-itemset RST candidates being generated by joining $(n-1)$–itemset frequent RSTs. Candidates in which one or more $(n-1)$-subsets are infrequent are pruned, for they cannot be frequent. Remaining candidates are then filtered with respect to support and confidence. Finally, frequent TPs are built by joining all RSTs in each frequent RST set.

Eventually, Stream Tree Miner (STMer) mines frequent labeled ordered subtrees over a tree-structured data stream \cite{HsiehWC06}. Its main contribution lies in candidate subtree generation, which is suitably incremental.

\subsection{Tree pattern rewriting}
\label{sec:tpqreformulation}

Query rewriting is casually used when views, whether they are materialized or not, are defined over data. Rewriting a query $Q$ that runs against a database $D$ is formulating a query $Q'$ that runs against a view $V$ built from $D$ such that $Q$ and $Q'$ output the same result. When data are tree-structured, views are predefined TPs. Then, $Q'$ is found among so-called useful embeddings (UEs) of $Q$ in $V$; the problem being to prune redundant, useless UEs \cite{WangWL09}. The heuristics that address this issue \cite{BalminOBCP04,LakshmananWZ06,WangWL09} rely on containment testing (Section~\ref{sec:EquivalenceAndContainmentTests}) to output a minimal set of UEs.

Query rewriting is also notably used in Peer Data Management Systems (PDMSs). In a PDMS, each peer is associated with a schema that represents the peer's domain of interest, as well as semantic relationships to neighboring peers. Thus, a query over one peer can obtain relevant data from any reachable peer in the PDMS. Semantic paths are traversed by reformulating queries at a peer into queries on its neighbors, and then to the neighbors' neighbors, recursively.
In the Piazza PDMS \cite{TatarinovH04}, data is modeled in XML, peer schemas in XML Schema, and queries in a subset of XQuery. Thus, query reformulation optimization strongly relates to TP matching optimization. For instance, one query may follow multiple paths in a PDMS, and thus may induce redundant reformulations, i.e., redundant queries on the peers. Pruning queries help reduce such redundancy. Pruning requires checking query containment (Section~\ref{sec:EquivalenceAndContainmentTests}) between a previously obtained reformulation and a new one. The reformulating process may also introduce redundant subexpressions in a query, leading to query minimization (Section~\ref{sec:PatternTreeMinimization}).

\subsection{Extended matching}
\label{sec:xmatching}

We review in this section two ways of ``pushing'' matching further on. The first one relates to checking the satisfiability of TPs against Active XML (AXML) documents \cite{MaHZ08,AbiteboulBM09}. AXML documents contain both data defined in extension (as in XML documents) and in intention, by means of Web service calls \cite{AbiteboulBMMW02}. When a Web service is invoked, its result is inserted into the document. In this context, \emph{a priori} checking whether a query (TP) is satisfiable, i.e., whether there exists any document the TP matches against, helps avoid unnecessary query computations.

Although TP satisfiability is well studied \cite{LakshmananRWZ04}, with AXML documents, Web service calls may also return data that contribute to query results, which makes the problem even more complex. Here, some fact is satisfiable for an AXML document and a query if it can be in the query result \emph{in some future state} \cite{AbiteboulBM09}. As Miklau and Suciu did for containment testing (Section~\ref{sec:EquivalenceAndContainmentTests}), Ma \emph{et al.} use tree automata to represent both TPs and sets of AXML documents conforming to a given schema. Then, the product tree automaton helps decide whether a TP matches any document. Abiteboul \emph{et al.} further test the relevance of Web service calls, i.e., whether the result can impact query answer \cite{AbiteboulBM09}.

The other example of matching enhancement lies in the context of building semantic communities, i.e., clusters of users with similar interests modeled as TPs. Chand \emph{et al.} propose to replace equivalence by similarity in matching, which could more generally be used to approximate XML queries \cite{ChandFG07}, as TP relaxation (Section~\ref{sec:TreePatternRelaxation}). Chand \emph{et al.} formulate the TP similarity problem as follows. Let $S$ be a set of TPs, $D$ a set of data trees, and $p, q \in S$. The similarity of $p$ and $q$ is a function $sim: S^2 \mapsto [0, 1]$ such that $sim(p, q)$ is the probability that $p$ matches the same subset of data from $D$ as $q$. Note that, depending on the proximity metric, $sim(p, q)$ may be different from $sim(q, p)$.

\section{Conclusion}
\label{conclusion}

\IEEEPARstart{W}{e} provide in this paper a comprehensive survey about XML tree patterns, which are nowadays considered crucial in XML querying and its optimization. We first compare TPs from a structural point of view, concluding that the richer a TP is with matching possibilities, the larger the subset of XQuery/XPath it encompasses, and thus the closer to user expectations it is.

Secondly, acknowledging that TP querying, i.e., matching a TP against a data tree, is central in TP usage, we review methods for TP matching optimization. They belong to two main families: TP minimization and holistic matching. We trust we provide a good overview of these approaches' evolution, and highlight the best algorithms in each family as of today. Moreover, we want to emphasize that TP minimization and holistic matching are complementary and should both be used to wholly optimize TP matching. 

We eventually illustrate how TPs are actually exploited in several application domains such as system optimization, network routing or knowledge discovery from XML sources. We especially demonstrate the use of frequent TP mining and TP rewriting for various purposes.

Although TP-related research, which has been ongoing for more than a decade, could look mature in the light of this survey, it is perpetually challenged by the ever-growing acceptance and usage of XML. For instance, recent applications require either querying data with a complex or only partially known structure, or integrating heterogeneous XML data sources (e.g., when dealing with streams). The keyword search-based languages that address these problems cannot be expressed with TPs \cite{TheodoratosW09}. Thus, TPs must be extended, e.g., by so-called partial tree-pattern queries (PTPQs) that allow the partial specification of a TP and are not restricted by a total order on nodes \cite{PlacekTSDS08,TheodoratosW09}. In turn, adapted matching procedures must be devised \cite{WuTSDS09}, a trend that is likely to perpetuate in the near future.

Moreover, we purposely focus on the core of TP-related topics in this survey (namely, TPs themselves, matching issues and a couple of applications). There is nonetheless a large number of important topics that we could not address due to space constraints, such as TP indexing, TP-based view selection, TP for probabilistic XML and continuous TP matching over XML streams. 


%



\ifCLASSOPTIONcompsoc
  \section*{Acknowledgments}
\else
  \section*{Acknowledgment}
\fi

We would like to thank all the anonymous reviewers of this paper for their thoughtful comments, and especially the reviewers of the very first version, who expressed interest in this work and encouraged us to write this widely complemented version.

We would also like to thank Mr. Huayu Wu, from the Institute of InfoComm Research, Singapore, for his valuable correspondence.


\ifCLASSOPTIONcaptionsoff
  \newpage
\fi



\bibliographystyle{IEEEtran}

\begin{IEEEbiography}[{\includegraphics[width=1in,height=1.25in,clip,keepaspectratio]{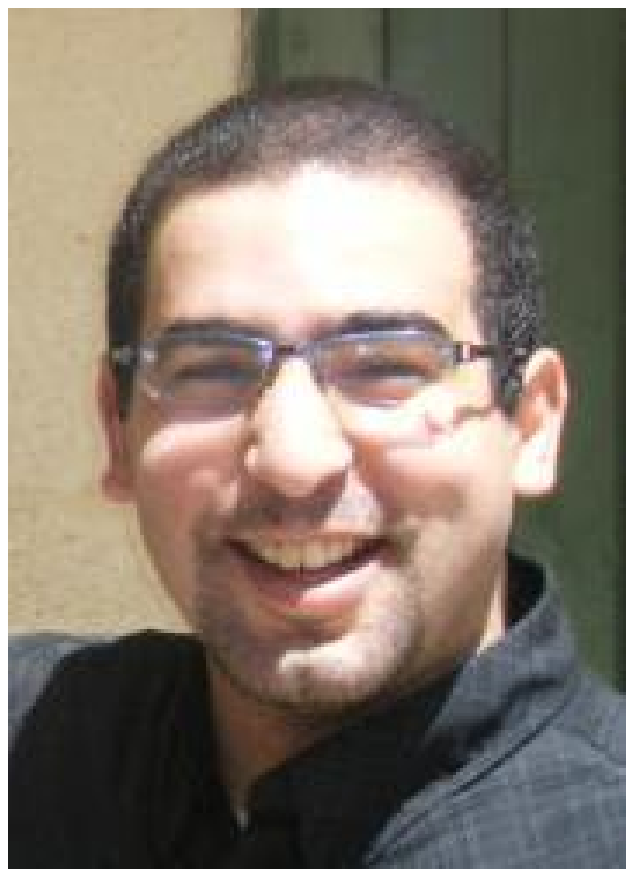}}]{Marouane Hachicha}
received his M.Sc. in computer science from the University of Lyon~2, France in 2007. He has been a Ph.D. student at the University of Lyon 2 since then. He spent six months as a foreign researcher in Italy in 2009. His research interests lie in XML data warehousing and XOLAP.
\end{IEEEbiography}

\begin{IEEEbiography}[{\includegraphics[width=1in,height=1.25in,clip,keepaspectratio]{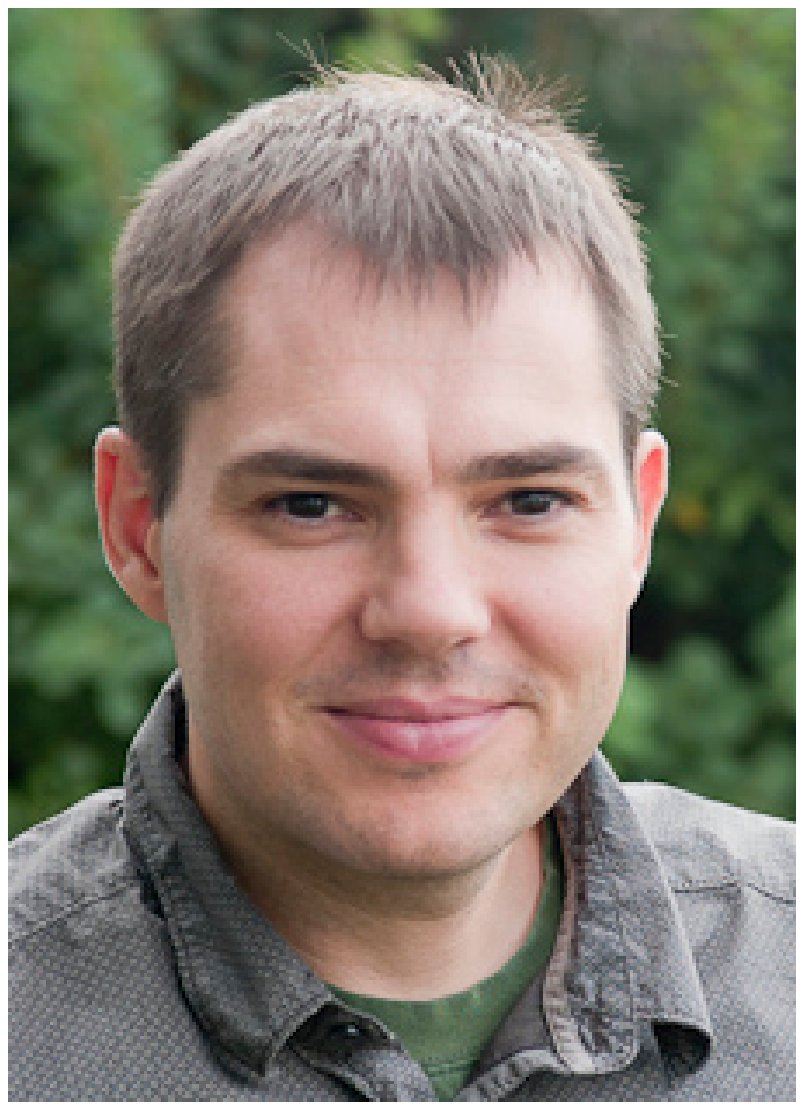}}]{J\'{e}r\^{o}me Darmont}
received his Ph.D. in computer science from the University of Clermont-Ferrand~II, France in 1999. He joined the University of Lyon 2, France in 1999 as an associate professor, and became full professor in 2008. He was head of the Decision Support Databases research group within the ERIC laboratory from 2000 to 2008, director of the Computer Science and Statistics Department of the Faculty of Economics and Management from 2003 to 2010, and has been in charge of the Complex Data Warehousing and OLAP research axis at ERIC since 2010. His current research interests mainly relate to handling so-called complex data in data warehouses (XML warehousing, performance optimization, auto-administration, benchmarking...), but also include data quality and security, cloud business intelligence, as well as medical or health-related applications.
\end{IEEEbiography}

\end{document}